\documentclass[conference,9pt]{IEEEtran}
\IEEEoverridecommandlockouts
\usepackage{verbatim}
\usepackage{float}
\usepackage{cite}
\usepackage{wrapfig}
\usepackage{graphicx}
\usepackage{amsmath,amssymb,amsfonts}
\usepackage{algorithm}
\usepackage[noend]{algpseudocode}
\usepackage{graphicx}
\usepackage{textcomp}
\usepackage{xcolor}
\def\BibTeX{{\rm B\kern-.05em{\sc i\kern-.025em b}\kern-.08em
    T\kern-.1667em\lower.7ex\hbox{E}\kern-.125emX}}
\usepackage[noend]{algpseudocode}
\usepackage{mathtools}    
\usepackage{amsmath}    
\usepackage{tikz}    
\usepackage{setspace}
\usepackage[font=small]{caption}
\usepackage[font=footnotesize]{subcaption}

\begin{document}
\captionsetup{font=small}
\title{A Learning Assisted Method for Uncovering Power Grid Generation and Distribution System Vulnerabilities \\
}

\author{Suman Maiti, Anjana B, Sunandan Adhikary, Ipsita Koley, Soumyajit Dey\\
Department of Computer Science and Engineering, Indian Institute of Technology, Kharagpur, India
}

\maketitle

\begin{abstract}
Intelligent attackers can suitably tamper sensor/actuator data at various  Smart grid surfaces causing intentional power oscillations, which if left undetected, can lead to voltage disruptions. We develop a novel combination of formal methods and machine learning tools that learns power system dynamics with the objective of  generating unsafe yet stealthy false data based  attack sequences. We enable the grid with  anomaly detectors in a generalized manner so that it is difficult for an attacker to remain undetected. Our methodology, when applied on an IEEE 14 bus power grid model, uncovers stealthy attack vectors even in presence of such detectors.
\end{abstract}

\section{Introduction}
\label{intro} 

Modern power grids implement software-based controls, smart sensing, actuation, and connectivity for efficient and sustainable operation. However, this also opens up  new  security  vulnerabilities  ~\cite{liang2016review}, like the coordinated cyber attack on the Ukrainian smart grid~\cite{liang20162015}. Typical attacks on power systems are launched by compromising the sensors, actuators, and communication mediums used for control and monitoring~\cite{mo2010false,amini2015dynamic,soltan2018blackiot}. \emph{False data injection attacks (FDIA)} and \emph{load alteration attacks (LAA)} are some primary examples of such attacks that involve falsification of the sensor measurements and abnormal alteration of the input loads connected to the grid. With a suitable choice of vulnerable points for data falsification and by smartly choosing the frequency of induced load alterations, a stealthy attacker may cause significant disturbances in power flow, known as \emph{unstable power swing}. This affects the generation and distribution components of a power grid~\cite{mo2010false,amini2015dynamic} leading to a loss in \emph{synchronism} among multiple generation units. Consequences are excessive brownouts or load shedding.  Moreover, such power demand manipulation attacks may also potentially manipulate stock markets  ~\cite{shekari2021mamiot}.

\par The effect of FDIA or LAA on the load inputs to  \emph{automatic generation control (AGC)} units in terms of inducing unstable power swings as have been widely studied in~\cite{deka2015one,he2020coordinated, tan2016optimal, anwar2016stealthy, law2014security}. In \cite{deka2015one}, the researchers discuss how manipulating a single circuit breaker ensures a successful attack on the power grid. The authors of  \cite{huang2018algorithmic} discuss an algorithmic approach for the synthesis of FDIA on power grid actuators in presence of the North-American electric reliability corporation’s critical infrastructure protection (NERC-CIP) standards for anomaly detection, which entirely ignores the transient dynamics of the system. The stealthiness of the attacks against quick response-specific detectors to detect the loss of synchronism between generators and the rest of the grid is not realized in either of these works. Authors in~\cite{tan2016optimal} generate fast and stealthy FDIA attacks on input loads of AGC by constraint solving, in the presence of bad data detectors. However such methodologies mostly scale for a single variable (say, the power flow) corresponding to a single attack surface. A similar detection approach is implemented in~\cite{he2020coordinated} to generate different types of attack co-ordinations on AGC while staying stealthy. But the focus of the work lies in coordination rather than synthesis. The work reported in \cite{jafari2021false} discusses launching of FDIA to  create \emph{false relay operations} (FRO) in the power grid by solving \emph{constraint satisfaction problems} (CSP) using satisfiability modulo theories (SMT). But the work mainly focuses on power grids containing  renewable energy resources. 

\par There also exists AI-based attack strategies that smartly generate stealthy yet effective FDIA or LAA on AGC systems. The work mentioned in \cite{guo2021reinforcement} has employed reinforcement learning (RL) to modify the electrical loads. The authors in~\cite{law2014security} have modeled the FDIA attack on power grids as a 2-player game with a security risk minimization objective. However, in these works, the attack sequences uncovered are not guaranteed to drive the system to an unsafe state. In \cite{marnerides2014power}, the authors have employed data analytics  to uncover anomalies in the power flow of the grid. In a similar vein, the authors in~\cite{anwar2016stealthy} propose blind FDIA by effectively utilizing partial system knowledge. These  approaches are statistically powerful but are  challenging to implement effectively~\cite{ahmed2020challenges} and they do not guarantee attack success.  

\par Evidently, the prominent contingency analysis-based detection methods either  largely ignore the transitory vulnerabilities of power systems for attack detection or their developments are model specific. Moreover, the learning-based and data-driven attack synthesis methods do not guarantee the stealth and success of the generated attack vectors. 
Drawing motivation from these gaps in state-of-the-art, we evolve a method for attack  vector discovery, which - 1) considers multiple attack surfaces together, and 2) considers the presence of generalised norm-based detection units that can analyse the transient dynamics of any power system model under attack. This provides a useful CAD flow for power system vulnerability  analysis which can uncover  {\em lethal and stealthy} attacks. 

\par\noindent\textbf{Contribution:} We develop an attack synthesis framework 
 that can take a   power grid model (currently Simulink based) as input along with their detection unit specifications and learn to generate the most effective LAA vectors and FDIA inputs that are guaranteed to maintain  stealth until the system is successfully made unsafe. The tool has the following novel components.

\par\noindent\textit{(i)} As a first phase, a  Reinforcement Learning (RL)-based attacker agent learns to inject the most effective input load altering attack sequences (LAAS) to push the system to its transient states for an elongated period as quickly as possible without raising any alarm.

\par\noindent\textit{(ii)} The probabilistic LAAS thus recovered are automatically augmented in the system model. This is given as input to a well-known simulation-based  falsification engine {\em S-TaLiRo} that synthesizes  false data sequences with the aim to falsify the set points of the generation unit. These additional FDIA perturbations  thus uncovered, are {\em formally guaranteed} to send the  generator units out of \emph{synchrony} with the rest of the power grid exploiting the transient characteristics induced by previously synthesized (phase \emph{(i)}) smart load alterations while maintaining stealth.

\par\noindent\textit{(iii)} We visualize the influence of the synthesized attacks in presence of a norm-based detection mechanism on an IEEE 14-bus power grid model by implementing it in a real-time hardware-in-loop (HIL) setup. The implemented norm-based detection mechanism is a generic representation of common anomaly detectors used in power grids.

\par Our attack synthesis methodology is a multivariate one, as it targets multiple attack surfaces of any input power grid model. 
The novelty of our methodology lies in partitioning the synthesis process among a learning-based probabilistic  engine and a stochastic optimization-based formal engine in order to uncover stealthy and lethal multivariate  attack sequences for complex grid systems in a scalable way. 

\section{System Modeling}\label{secSysDesc}
\begin{figure}[H]
\includegraphics[width = \columnwidth]{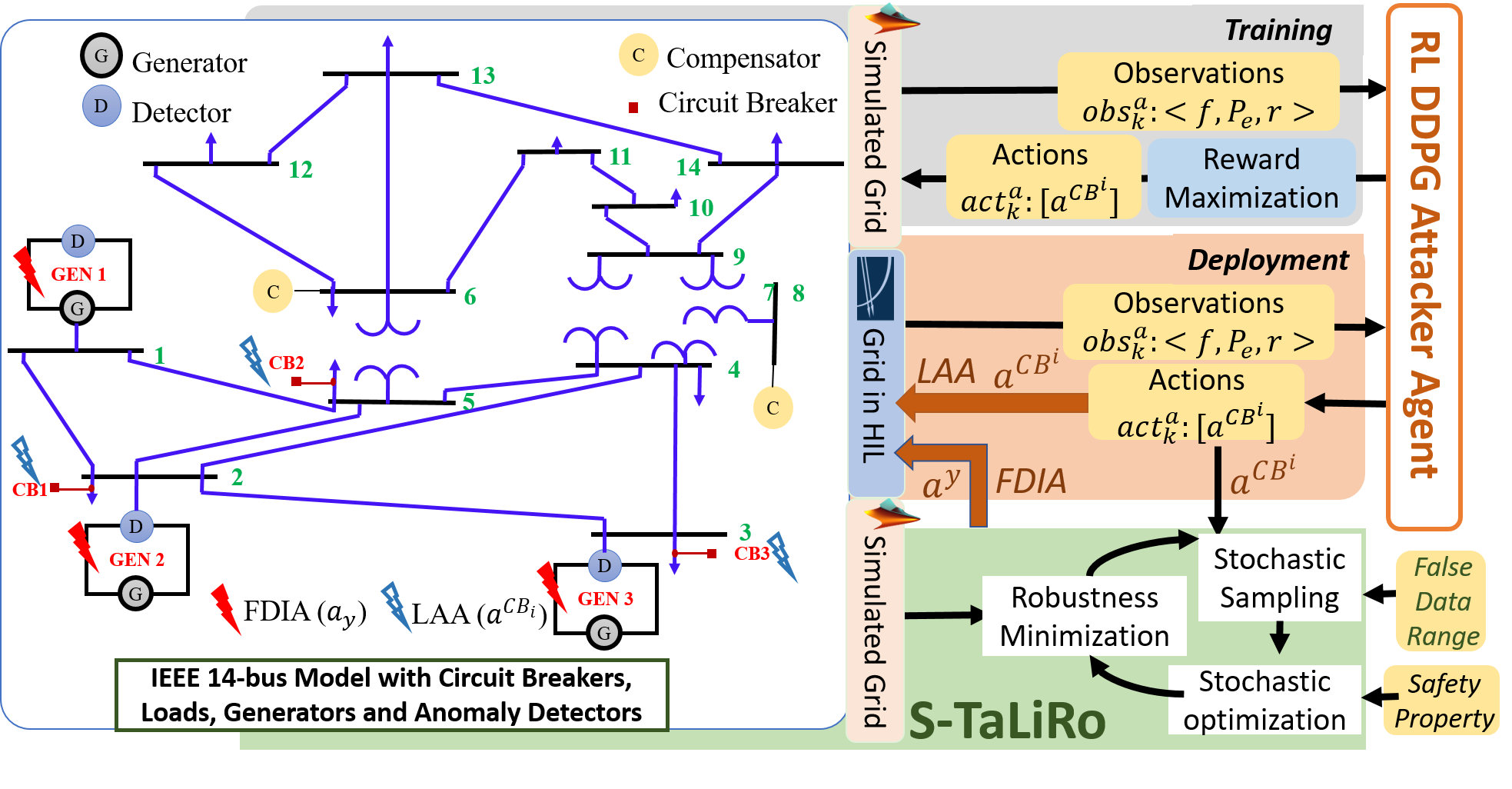}
\centering
\caption{Tool setup used for the attack generation.}
\label{fig:figtool}
\end{figure}
\begin{figure}[H]
\centering
\includegraphics[width = 0.95\columnwidth]{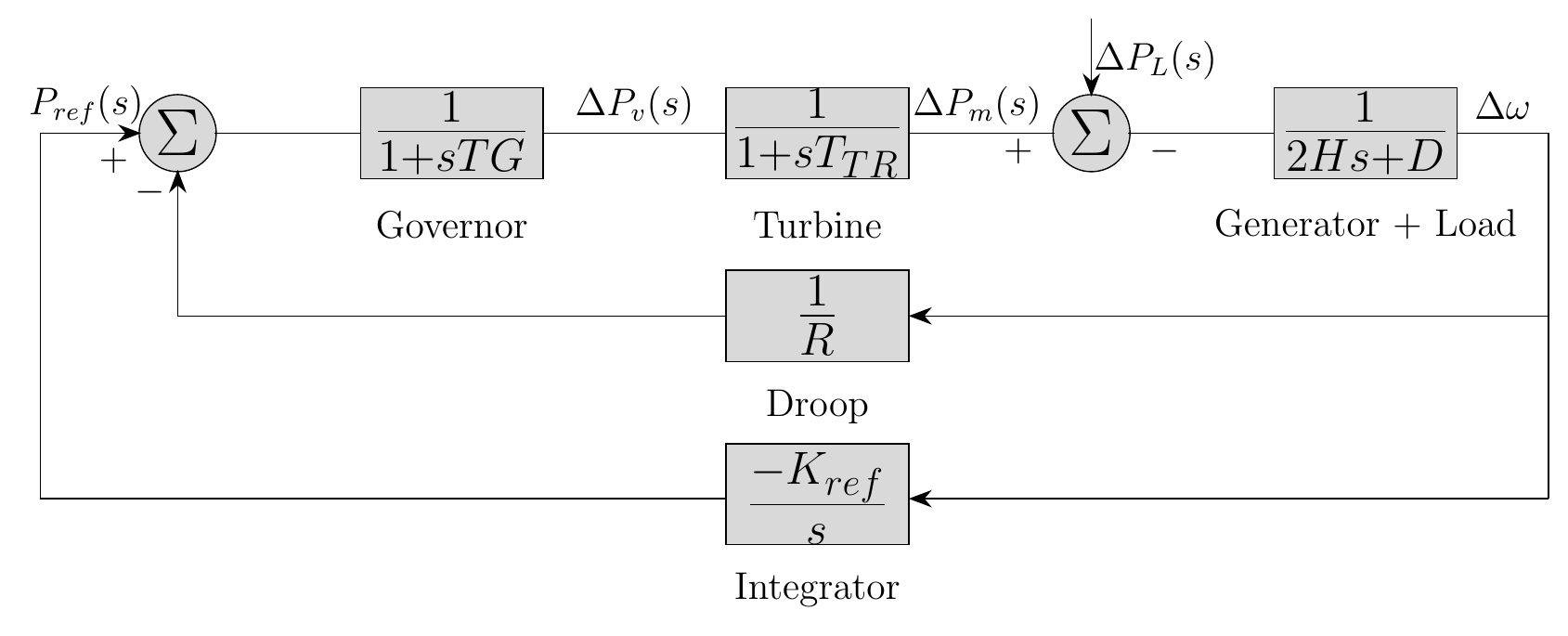}
\centering
	\caption{AGC Closed-Loop for A Synchronous Generator~\cite{saadat1999power}}
	\label{fig:figagc1}
\end{figure}

For identifying attack surfaces, let us consider an IEEE 14 bus power grid  model (refer to Fig.\ref{fig:figtool})   having generation units that include  \emph{three synchronous generators}, each with a nominal power rating of 100 Mega Watts (MW)  and nominal voltage rating of 146.280 Kilo Volts (KV). The transmission units consist of transmission lines and 14 buses. 
 Among the 14 buses, $one$ bus is a slack bus (bus-1), which serves as an angular reference for all other buses in the grid; $three$ are PV buses connected with the generators to handle constant power and voltage generation, and the rest are PQ Buses used to find the voltage and phase angle of all the buses. 
  Generation units employ the AGC strategy to meet the change in load demand and eventually retain the grid frequency to its nominal value of $60$Hz. AGC consists of four components: \textit{(i)} generator unit, \textit{(ii)} turbine unit, \textit{(iii)} speed governor system, and \textit{(iv)} the electrical loads (refer Fig.\ref{fig:figagc1}). For the generator to supply additional power to the grid, the mechanical power ($P_m$) supplied by the turbine should be changed by adjusting the amount of steam flowing into the turbine, which in turn is controlled by the speed governor. The droop ($D=1/R$) serves as a constant feedback gain for the governor system and calculates the generated power from the generator rotor speed. Thus the speed governor keeps track of the generated power and decides how much speed is required to meet the power demand or the reference set point-power ($P_{ref}$). The integrator is used as another feedback unit for updating this reference set point-power ($P_{ref}$) of the governor to meet the changes in the load demand (refer \cite{saadat1999power}).
We design the following continuous-time state-space model of the AGC-synchronous generator with the symbols listed in Table~\ref{tab:params}: 

{\small
\begin{align}
\label{eqsscont}
&\begin{bmatrix}\dot{\Delta\omega} \\ \dot{\Delta P_{m}} \\ \dot{\Delta P_{v}} \\ \dot{\Delta P_{ref}} \end{bmatrix} =\begin{bmatrix}\frac{-D}{2H} & \frac{1}{2H} & 0 & 0 \\ 0 & \frac{-1}{T_{TR}} & \frac{1}{T_{TR}} & 0 \\ \frac{1}{RT_{G}} & 0 & \frac{-1}{T_{G}} & \frac{1}{T_{G}} \\ -K_{ref} & 0 & 0 & 0\end{bmatrix}  {\begin{bmatrix}\Delta\omega \\ \Delta P_{m} \\ \Delta P_{v} \\ \Delta P_{ref} \end{bmatrix}} + {\begin{bmatrix} \frac{-1}{2H} \\ 0 \\ 0 \\ 0 \end{bmatrix}} \Delta P_{L}\\
&\begin{bmatrix} \Delta\omega & \Delta P_{ref} \end{bmatrix}^T = \begin{bmatrix} 1 &0&0&0\\0&0&0&1 \end{bmatrix} 
\begin{bmatrix}
\Delta\omega&\Delta P_{m}&\Delta P_{v}& \Delta P_{ref}
\end{bmatrix}^T
\nonumber
\end{align}
}
Here state vector $x = [\Delta\omega\ \Delta P_{m}\ \Delta P_{v}\ \Delta P_{ref} ]^{T}$, input vector $u = [\Delta P_{L}]$ and output vector $y=[\Delta \omega \ \Delta P_{ref}]^T$. Note that, the load input is user controlled. We consider the discrete-time version of this linear time-invariant (LTI) AGC closed-loop as our target system: 
\hspace{1mm}
{\small
\begin{align}
\label{eqa22}
    &x_{k+1} = Ax_{k} + Bu_{k};\ y_{k} = Cx_{k} + Du_{k}\\\nonumber
    &{\hat{x}_{k+1}} = (A - LC) \hat{x}_{k} + B u_{k} + L y_{k} ;\ r_{k} = y_{k} - \hat{y}_{k} = y_{k} - C \hat{x}_{k}
\end{align}
}
Here $A,B,C,D$ are the discretized version of the transfer matrices from Eq.~\ref{eqsscont}, $K$ is the feedback controller gain and $L$ is the Kalman gain. We use a Kalman filter-based estimator here in order to estimate the states from the system output and calculate the control action accordingly. $\hat{x}$, $\hat{y}$, and $r_{k}$ are the estimated state, estimated output, and residue of the discretized AGC closed-loop respectively. All variables are subscripted with the corresponding sampling iteration, e.g. $x_k$ denotes the state values at $k$-th sampling period.

\begin{wraptable}[9]{l}{0.55\columnwidth}
 \scriptsize
\begin{tabular}{|l|l|}
\hline
\multicolumn{1}{|c|}{Symbols} & \multicolumn{1}{c|}{Descriptions} \\ \hline
$\Delta\omega$ & Change in rotor speed \\ \hline
$\Delta P_m$ & Change in Mechanical Power \\ \hline
$\Delta P_v$ & Change in Valve Position \\ \hline
$\Delta P_{L}$ & Change in electrical load \\ \hline
$D=1/R$ & Droop \\ \hline
$H$ & Inertia Constant \\ \hline
$T_{TR}$ & Transmission Time Delay \\ \hline
$T_G$ & Generation Time Delay \\ \hline
$K_{ref}$ & Feedback Gain \\ \hline
\end{tabular}
\vspace{-2 mm}
\caption{List of Symbols}
\label{tab:params}
\end{wraptable}
\par \textbf{Anomaly Detector:}
The Kalman gain $L$ is designed to filter out  process and measurement noises and optimally reduce system  residue i.e. the difference between actual and  estimated output. We use a residue-based detector as an anomaly detection unit. It checks whether under nominal behavior the $\infty$-norm of the system residue remains within certain pre-decided threshold ($Th$), i.e. $\max\limits_{i=0}^{m-1}|r_i|\leq Th$. Violation of this criteria raises suspicion about anomalous system signals since the actual and estimated outputs differ significantly. Such a residue-based detection generalises the bad-data detection units that are usually present in the power grid for anomaly detection. Also, they can capture the transient behaviour of the system by tracking the residue. In the next section, we discuss how we plan to launch attacks on an AGC loop in presence of such an anomaly detector.

\section{Proposed Methodology}\label{secMeth}
\subsection{Attack Modeling}
\label{secAtkMod}
Some accessible components of the power grid are normally used by the power operators for monitoring, sensing, and control to ensure the safe operation of the power grid. An attacker can exploit the vulnerable exposure of these accessible components and launch desired alterations or falsifications of control or output signals. 
\subsubsection{Choosing Vulnerable Points} One such vulnerable point can be the \emph{circuit breakers} which are the physical devices controlled by computer-enabled relays. They connect/disconnect a power system component such as a generator or a load to/from the rest of the grid. A suitably crafted connection-disconnection sequence can therefore alter the loads connected to the power grid thus changing the demand. A frequent enough load-altering sequence can significantly manipulate the  rate of change in the frequency-independent power load $\Delta P_{L}$ for the grid. Consider there are $m$ circuit breakers and the control signal of the $i$-th circuit breaker is $a^{CB_i}$. An attacker alters the control signals of the circuit breakers for $d$ sampling iterations, i.e. [$a^{CB_i}_1,a^{CB_i}_2,\cdots a^{CB_i}_d, \forall i\in [1,m]$] are under attack. This induces an abrupt change in frequency-independent load $\Delta P_{L_k}=h(a^{CB_i}_k,a^{CB_i}_k,\cdots a^{CB_i}_k, \forall i\in [1,m])$. Here,  $h()$ is a function that maps the change in $m$ circuit breaker control inputs to the change in load input $\Delta P_L$. Typically in such a load alteration attack (LAA)~\cite{deka2015one,law2014security}, the goal of the attacker is to push the AGC system to its transient phase for a longer period in order to ensure that the generators are out of $synchronism$ while remaining stealthy.
\par Another such vulnerable point of attack is the output signals from the generator. 
As explained in Sec.~\ref{secSysDesc}, the AGC is implemented with every generation unit to sense the change in desired power reference $\Delta P_{ref}$ based on the change in grid frequency $\Delta \omega$. The feedback loop of AGC accordingly updates the $P_{ref}$ which is input to the speed governor. Based on this, the governor decides how much mechanical power is to be actuated to the steam valve so that the turbine rotates to generate enough torque. This extra torque is fed to the generator so that the desired power demand is met. A false data injection attacker can change this $\Delta P_{ref}$ by adding false data $a^y$ for consecutive $d$ sampling iterations such that $\Delta \tilde{P}^a_{ref}=\Delta P_{ref}+a^y$ is fed back to the governor in Eq.~\ref{eqsscont}. This makes the speed governor unaware of the actual required change in the desired power, which is to be generated for the same time window.

\subsubsection{Motivation and Intuition behind combining LAA and FDIA}
\begin{wrapfigure}{l}{0.5\columnwidth}
\centering
\begin{subfigure}[b]{0.53\columnwidth}
   \includegraphics[clip,width=\columnwidth]{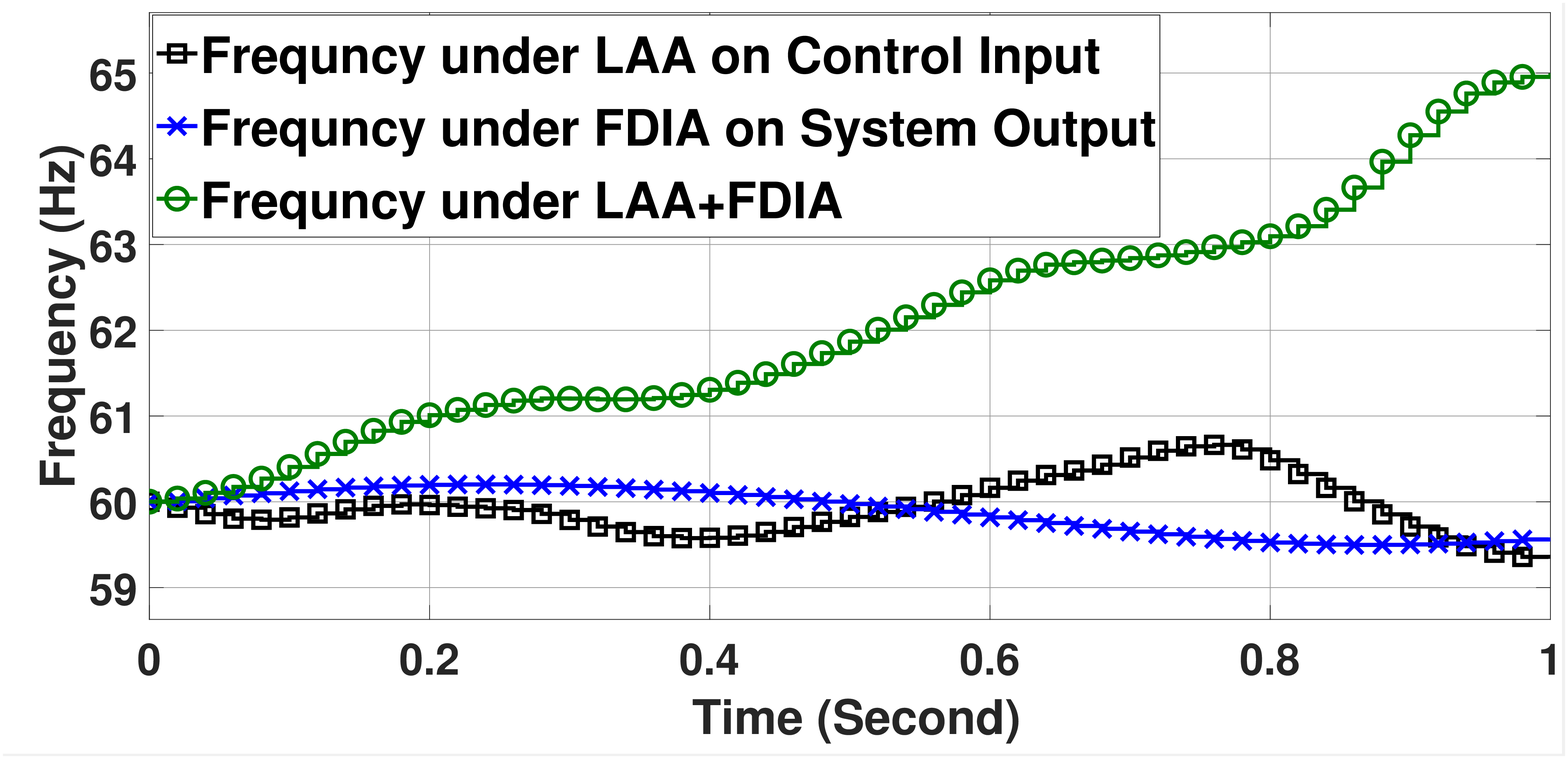}
\caption{Operating frequency comparison of generator $GEN1$ under attack simulated in simulink}
\label{fig:figcomp} 
\end{subfigure}
\\
\begin{subfigure}[b]{0.53\columnwidth}
    \includegraphics[clip,width=\columnwidth]{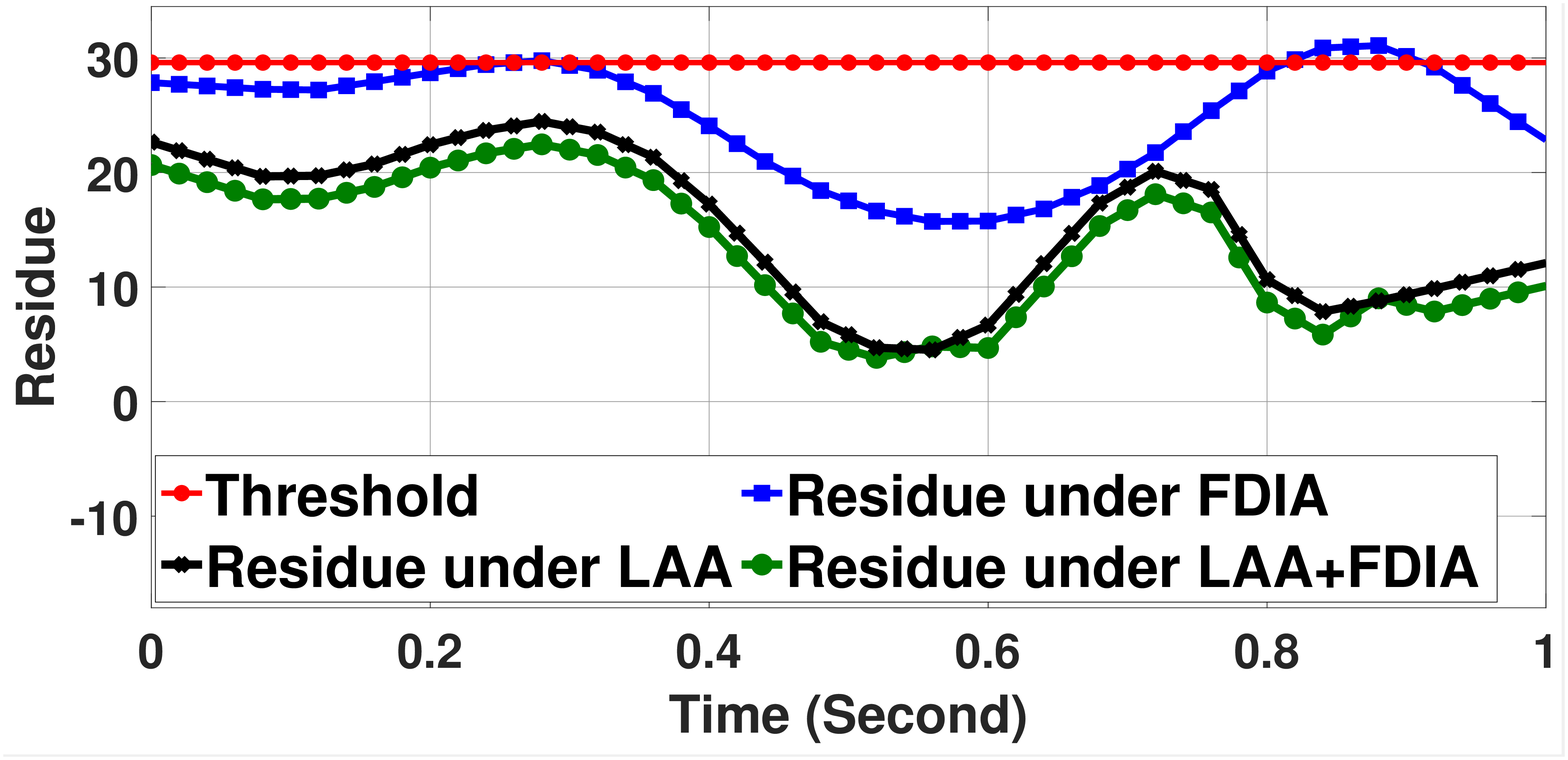}
\caption{Residue comparison of $GEN1$ under attack simulated in simulink}
\label{fig:figcompres1}
\end{subfigure}
\end{wrapfigure}

We demonstrate how the LAA and FDIA affect an IEEE 14-bus power grid model by plotting frequency and residue under attack in Fig.~\ref{fig:figcomp} and Fig.~\ref{fig:figcompres1}. 
As we can see, for an optimal LAA sequence, the frequency ($black$ plot in Fig.~\ref{fig:figcomp}) of the generator deviated from its safety range ($60\pm 0.5$) for a small fraction of time. The attack remained stealthy (i.e. less than the threshold, $red$ plot in Fig.\ref{fig:figcompres1}) for a period of more than 1 second (see Fig.\ref{fig:figcompres1}). Therefore, the LAA is successful in stealthily inducing a certain amount of transient behaviour in the system but for a short duration. Whereas, the FDIA got detected at $0.3$ second (see Fig.\ref{fig:figcompres1}), while the frequency of the generator unit deviated significantly from their safety range but, only for a small fraction of time (refer Fig.\ref{fig:figcomp}). In this case, the FDIA had to be continued for a significant amount of time even after being detected, for the frequency to migrate outside the permissible range. To summarize, in none of the cases, the system remained unsafe for a long duration of operation with the attacks remaining stealthy.  

\par Now, we apply both these attack vectors together. We plot the change in frequency and residue of $GEN1$ under the combination of LAA and FDIA ($green$ plots) in Fig.~\ref{fig:figcomp} and Fig.~\ref{fig:figcompres1}. This causes a significant and fast change in frequency of $GEN1$, ($green$ plot in Fig.~\ref{fig:figcomp}), without the residue ($green$ plot in Fig.\ref{fig:figcompres1}) crossing the threshold ($red$ plot in Fig.\ref{fig:figcompres1}). We observed similar effects for several such LAA and FDIA combinations. Now, \emph{why this combination of LAA and FDIA is more effective}? This can be intuitively explained. 
Considering an LAA on the system load input ($\Delta P_L$) and FDIA on the reference power output ($\Delta P_{ref}$), the system equations in Eq.~\ref{eqa22} are modified as follows.

{\small
\begin{align}\label{eq:StateAtk}
    &x^a_{k+1} = A x^a_{k} + B \tilde{u}^a_k,\
    {y}^a_{k} = C x^a_{k},\ {r}^a_{k+1} 
    = y^a_{k+1} - C \hat{x}^a_{k+1} &\\\label{eq:EstAtk}
    &\hat{x}^a_{k+1} = A \hat{x}^a_{k} + B u^a_{k} + L r^a_{k},\ u^a_{k+1} = K\hat{x}^a_{k+1}
    &\\\label{eq:IOAtk}
    &\tilde{y}^a_{k+1} = {y}^a_{k+1} + a^{y}_{k+1},\ \tilde{u}^a_{k+1} = h(a^{CB_1}_{k+1},\cdots a^{CB_m}_{k+1})&
\end{align}
}
\hspace{5mm}
Here, $a^{y}_{k}$ is the FDIA injected on the output measurements, and $a^{CB_i}_{k+1}$ is the control signal for $i$-th circuit breaker injected as part of LAA at $k$-th sampling iteration.
$\tilde{u}^a_{k}$ is the altered load input due to the manipulation of circuit breaker signals at $k$-th time stamp. $\tilde{y}^a_{k}$ is the falsified output signal at $k$-th time stamp. All the usual system variables are superscripted with $a$ to denote their attacked variants. Under normal operation, the AGC closed-loop system progresses following Eq.~\ref{eqa22}. It is designed to  handle the change in the input loads and update the reference power input accordingly based on the change in the frequency (see Eq.~\ref{eqsscont}). Unless we introduce severely abnormal changes in the circuit breaker signals ($a^{CB_i}$), the system states ($x^a$) recover from transient to steady-state very quickly. This is because the estimator can track the actual states since the outputs are not falsified (estimator uses $y^a$ instead of $\tilde{y}^a$ in Eq.~\ref{eq:EstAtk}). 
If a residue-based detection unit is present to flag any anomalous difference between actual and estimated outputs ($||r^a||_\infty$), the LAA cannot modify the control input significantly. As a result, even with the altered input loads, the system state recovers back to safety. Whereas, if we falsify the required change in reference power while inducing the maximum possible stealthy load alterations, the AGC unit does not get to see the actual power demand to be met and tracks a wrong output following Eq.~\ref{eq:EstAtk}.
Hence, there always remains a difference between the generated electrical power and the electrical load, i.e. actual state under attack ($x^a$) and the estimated state under attack ($\hat{x}^a$). This falsified sensor measurement ($\tilde{y}^a$) blinds the detection unit from detecting the heavy input load alterations ($\tilde{u}^a$). This causes the frequency to change (increase or decrease based on the false data) in a way such that it does not flag the detector but a cascading effect of this change pushes the generation units out of synchrony. 
\subsection{Problem Formulation}\label{secProb}
Based on the earlier discussion, the goal of an attacker is to inject false data and alter the control inputs simultaneously in order to push the system outputs beyond their safety bounds while the anomaly detector is not triggerred~\cite{teixeira2015secure}. An attacker is considered $(i)$ $successful$  when it pushes the system outputs (i.e. frequency of the generator unit, which is proportional to $\omega$) beyond its safety range i.e., ($60\pm 0.5$) of the  nominal frequency (refer to \cite{obaid2019frequency}) and $(ii)$ $stealthy$ when keeps the maximum system residue below a certain threshold. We define an attack vector $A_{d} = \begin{bmatrix}
\tilde{u}^a_{k_{0}},\tilde{u}^a_{k_{0}+1},...,\tilde{u}^a_{k_{0}+d}\\
a^y_{k_{0}},a^y_{k_{0}+1},...,a^y_{k_{0}+d}\end{bmatrix}$, where $a^y_{k}$ denotes the false-data injected to system outputs and $\tilde{u}^a_{k}$ denotes the manipulated load inputs to the system at each sampling iteration $k\leq d$ (see Eq.~\ref{eq:StateAtk}-\ref{eq:IOAtk}). In this work, our goal is \emph{to synthesize load alteration and false data injection sequences for power grid such that the following criteria holds:}
\begin{align}
    \exists k \geq0\ s.t.\ \tilde{y}^{a}_{k^\prime} \notin {Safe}_{y} \text{ and } 
    {\| r_k \|}_{\infty} \leq {Th}\ ;\ \forall k < k^ \prime
    \label{eqa2}
\end{align}
Informally, this means we intend to synthesize an attack vector $A_d$ such that there exists a sampling iteration $k^\prime>k\geq 0$, for which the falsified system output $\tilde{y}^a_{k^\prime}$ violates safety envelope ${Safe}_{y}$ while the infinity norm of residue under attack (${\| r_k \|}_{\infty}$) remains within the certain predefined threshold ($Th$) before that. 
In the following section, we explain our novel methodology to synthesize such successful and stealthy combinations of input LAA and FDIA on outputs.


\subsection{Attack Synthesis}\label{secAtkSynth}
In this work, the proposed strategy generates formally verified combinations of FDIA and LAA sequences to make any power grid model unsafe. The framework takes the following inputs: $(i)$ a power system model, $(ii)$ the detection unit specifications ($Th$ in Eq.~\ref{eqa2}), and $(iii)$ the range of false data to be synthesized. 
The framework outputs a $d$ length vector $A_d$ that contains a $d$ sampling iteration long optimal LAA 
 and a $d$-length FDIA. 
\subsubsection{ LAA Synthesis Using RL agent}\label{secAtkRl}
We propose a learning-based methodology to learn the most effective input load alteration sequences that $(i)$ keep the system residue under a certain threshold and $(ii)$ drive the system outputs (frequency) beyond the safe boundary as quickly as possible. Existing works~\cite{anwar2016stealthy,asadi2022data} demonstrate that data-driven modeling of a power system is possible. Following Sec.~\ref{secAtkMod}, we assume that the attacker can acquire full topological knowledge of the power grid using such data-driven approaches and has an access to the computer program that engages or disengages the available

circuit breakers ($CB1$, $CB2$ and $CB3$ in Fig.~\ref{fig:figtool}). 

We design an attacker RL Agent $\Lambda^a$ that  intelligently manipulates the input loads ($\tilde{u}^a$) by changing the circuit breaker control signals ($a^{CB_i}$) while observing the operating frequency $f$ and output power ($P_e$) of all generator units along with their system residues ($r$). Therefore, we define the actions of $\Lambda^a$ as ${act^a} = [a^{CB^j}\ \forall 1\leq j \leq m]$ considering there are $m$ circuit breakers (see Eq.~\ref{eq:IOAtk}). 

At every $k$-th step of an episode, the RL agent needs to observe $\{f^i, r^i, {P_{e}}^i\ \forall 1\leq i \leq n\}$, where $f^i$, ${P_{e}}^i$ indicate the frequency, electrical power of $i^{th}$ generator respectively and $r^i$ indicates the residue of the $i^{th}$ anomaly detector, considering there are $n$ generators. Note that, to observe the frequency of $i$-th generator unit,  the agent can observe its rotor speed $\omega^{i}$ i.e. output of the discussed state space model in Eq.~\ref{eqsscont} (since $f^{i} = \omega^{i}/2\pi$). 
 The RL agent also needs to observe its last set of actions in order to map the actions to the rewards.  Therefore, ${{obs^a}_{k}^i}$ = [${f_{k}^i}$,  ${r_{k}^i}$, ${{P_{e,k}}^{i}}$ ${act^a}_{k-1}$] at $k^{th}$ step of an episode. Considering these observations at every $k$-th simulation instance, the attacker agent aims to choose an action tuple  ${act^a_{k}}$ in order to maximize the reward $rwd^a_{k}  = [ w1  \times (\sum\limits_{i=0}^{n}\Call{check}{|{P_{e,k}^i}|\notin Safe_{P_{e}}}  \times \sum\limits_{i=0}^{n}\Call{check}{|r^i_k|\leq Th})  +  w2 \times (\sum\limits_{i=0}^{n}\Call{check}{f^i_{k}\notin Safe_{f}}  \times \Call{check}{|r^i_k|\leq Th})  +  w3 \sum\limits_{i=0}^{n}\times \Call{check}{|r^i_k|\leq Th}]$.  Here, $w1$, $w2$, and $w3$ are real-valued weights that denote the relative priorities of the reward components (refer~\cite{koley2021catch}). $\Call{check}{}$ represents a function that maps the predicate argument to 0/1 based on truth/falsify. We denote the safe regions of $f, P_e$ using the $Safe_{f}, Safe_{P_e}$ notations. The intuition behind creating such a reward function is to reward the attacker agent only if the system parameters i.e. $f$, $P_e$ are outside their safe range and the residue $r$ is below the threshold. Note that the reward is highest if the $act^a_{k}$ is chosen in such a way that all generators have their residues below the threshold and the frequency and electrical power is unsafe. Therefore, we can say the reward function $rwd^a_{k}$($obs^a_{k}$, $act^a_{k}$) quantifies the measurement of stealth and success of the chosen LAA. 

\par We have implemented the lerning framework using  \emph{Deep Deterministic Policy Gradient (DDPG)} algorithm \cite{lillicrap2019continuous}. 

The $actor$ neural network present in a DDPG agent deterministically chooses an action ($act^a$) by observing the states ($obs^a$) from the environment. A $critic$ deep Q-Network further analyzes the Q values generated for the last few actions taken by the $actor$ and their corresponding rewards for some observations. Then it criticizes the $actor$'s policy with the goal of maximization of reward. Required model-specific customizations over the standard DDPG algorithm (Algo.~1 in~\cite{lillicrap2019continuous}) are done so that the RL agent observes and learns the actions that are vulnerable to our system environment. A trained RL agent can then be deployed online as an attacker agent to launch LAA. 

\par As we have seen earlier in Sec.~\ref{secAtkMod}, LAA alone is not much effective on power grids. But LAA invokes transient characteristics of the power grid, which makes it stochastically promising to discover a simultaneously active false data injection strategy for guaranteed falsification of system safety while being stealthy all through the attack execution. We explain this next step of our methodology in the following section.

\subsubsection{FDIA Synthesis and  Validation}\label{secFormalAtkSyn} 
As part of our automation, the grid model is  automatically augmented with the LAA  sequences identified for the circuit breaker control signals and given as input to the second step of our methodology. In this step, we employ a well-known \emph{stochastic falsification}-based verification method \cite{abbas2013probabilistic} that searches for false data sequences  [$a^y_{k_0}\cdots a^y_{k+d}$]  such that the criteria in Eq.~\ref{eqa2} holds. The learning-based LAA strategy can only promise  probabilistic success in choosing the proper attack action. However, the FDIA synthesized in this step  validates the RL-generated LAA when applied in conjunction as the combined attack is guaranteed to steer the system to  unsafe states while maintaining a residue below the threshold (thus implying stealth). 

Algo.~\ref{alg:synnval} captures this attack synthesis and validation strategy. The algorithm takes the following inputs: $(i)$  the system model parameters, i.e. $\langle A^i,B^i,C^i,K^i,L^i,\,  \forall i \in [1,n], n \rangle$ ($n$ is the number of generators connected to the power grid), $(ii)$ the threshold $Th^i\ \forall i \in [1,n]$ for all detection units attached to the generators, $(iii)$ for each state variable $x$, the safety boundary $Safe_x$, $(iv)$ 
the optimal LAA [$a^{CB_j}_{k_0}\cdots a^{CB_j}_{k+d},\ \forall j\leq m$] 
generated by the RL agent and corresponding initial values of system variables $init^i\ \forall i \in [1,n]$, $(v)$ the length of desired FDIA sequence $d$ (same as the LAA sequence) $(vi)$ a closed range of false data values $Range_{a^y}$ and $(vii)$ attack mask $C_{atk}$ that is a $1\times q$ (considering the system as $q$ outputs) matrix that denotes, which output variables are falsified, i.e. $C_{atk}[i]=1$ means FDIA is on the $i$-th output.

\begin{algorithm}[!ht]
\footnotesize
\caption{Attack Vector Synthesis and Validation}
\label{alg:synnval}
\begin{algorithmic}[1]
\Require{$\langle A^i,B^i,C^i,K^i,L^i \forall i\in[1,n], n \rangle$, $Th^i$$\forall i\in[1,n]$, $Safe_{y}$, $\langle [a^{CB_j}_{k_0}\cdots a^{CB_j}_{k+d},\ \forall j\leq m],init^i \forall i \in[1,n]\rangle $, $d$, $Range_{a^y}$, $C_{atk}$}
\For { $i \in [1,n]$ }\label{algsyn:forstart}
\State $\hat x^{a,i}[0], x^{a,i}[0], u^{a,i}[0] \gets init^i, {y}^{a,i}[0] \gets Cx^a[0], r^{a,i}[0] \gets {y}^{a,i}[0] - C\hat{x}^{a,i}[0]$,$k\gets 1$,$[a^{y^i}[0]\cdots a^{y^i}[d]]\gets [0\cdots 0]$\Comment{initialisation}\label{algsyn:init}
\While {$k \leq d$}\label{algsyn:whilestart}
\State $\tilde{u}^{a,i}[k-1]\gets h(a^{CB_i}_{k-1} \forall i\in [1,m]))$\label{algsyn:laa}\Comment{Modeling LAA}
\State $x^{a,i}[k] \gets A^ix^{a,i}[k-1] + B^i\tilde{u}^{a,i}[k-1]$;\label{algsyn:x}
\State $\hat{x}^{a,i}[k] \gets A^i \hat{x}^{a,i}[k-1] + B^i u^{a,i}[k-1] + L^ir^{a,i}[k-1]$;\label{algsyn:xhat}
\State $u^{a,i}[k] \gets K^i\hat{x}^{a,i}[k]$;\label{algsyn:uatk}
\For {$j\in [1,q]$}\label{algsyn:formask}
\If {$C_{atk}[i]$ is  $1$} \label{algsyn:checkmask}
    \State $a^{y^i}[k] \gets chooseFromRange(Range_{a^y})$;
    \Comment{falsifying assignment}
    \label{algsyn:fdi}
\EndIf
\EndFor
\State ${y}^{a,i}[k] \gets C^i x^{a,i}[k] + a^{y^i}[k]$; \label{algsyn:yatk}
\State $r[k] \gets {y}^a[k] + C^i\hat{x}^{a,i}[k]$;
\State $k \gets k + 1$;\label{algsyn:r}
\EndWhile
\State $\phi^i \gets assert\neg((\|r^{a,i}[1]\|_{\infty} \leq Th)  \land...\land (\|r^{a,i}[d]\|_{\infty} $\\$\leq Th) \land (y^{a,i}[1]\notin Safe_{y})\vee \cdots \vee (y^{a,i}[d] \notin Safe_{y}))$;\label{algsyn:phii}
\EndFor
\State $\phi \gets {assert(\phi^1 \vee \cdots \vee \phi^n)}$; \label{algsyn:phi}
\If {$\phi$ is unsatisfiable} $a^{y^i}[k] \gets NULL$;\Comment{no counter-example}\label{algsyn:ifunsat}
\EndIf
\State \Return $A_{d} \gets \begin{bmatrix}
\tilde{u}_{1},\cdots,\tilde{u}_{d}\\
a^{y^i}[1],\cdots,a^{y^i}[d]\end{bmatrix}$\label{algsyn:ret}
\end{algorithmic}
\end{algorithm}

The algorithm first initialises AGC system variables for each generation unit with the given initial values or zeroes (line \ref{algsyn:init}). The input circuit breaker signals, generated by the RL for LAA are next  used to compute the attacks on control signals $\tilde{u}^a$ as part of the system model (line~\ref{algsyn:laa}). The system is then incorporated    with the altered load and false data inputs for $d$ sampling iterations following  Eq.~\ref{eq:StateAtk}-\ref{eq:IOAtk} (lines\ref{algsyn:laa}-\ref{algsyn:r}). We consider a new array of variable $a^{y^i}[k]$ representing the  false data to be injected at each $k$-th iteration to the output variable of the $i$-th generator $y^{a,i}[k]$ (see line \ref{algsyn:yatk}). We initialise them with $0$ (in line \ref{algsyn:init}) and specify their value range $Range_{a^y}$ using the  $chooseFromRange()$ function in every iteration according to the attack mask $C_{atk}$ (i.e. only for those outputs that we intend to attack, see line~\ref{algsyn:formask}-\ref{algsyn:fdi}). After unrolling the closed loop for $d$ iterations, we formulate the assertion $\phi_i$ that ensures the safety of the system i.e. negation of the $successful$ and $stealthy$ attack criteria from Eq.~\ref{eqa2} (line \ref{algsyn:phii}). The same is repeated for all the generation units (from line \ref{algsyn:forstart}-\ref{algsyn:phii}) and  by $or$-ing all the $\phi^i$-s, we build the final assertion $\phi$ to verify (line \ref{algsyn:phi}). We give this assertion as input to the simulation-based verification tool S-TaLiRo. S-TaLiRo looks for certain assignments to the non-deterministic variables ($a^{y^i}[0]...a^{y^i}[d]$ in our case) within a given range ($Range_{a^y}$) with help of stochastic optimization. It uses the Monte-Carlo sampling strategy to sample a possible assignment from the given range of these variables in order to look for a simulated output that is locally least robust w.r.t. the input assertion. The $chooseFromRange()$ function denotes this assignment to $a^{y^i}[k]$ (line~\ref{algsyn:fdi}). A counter-example trace that falsifies the input property is therefore the desired false data sequence for us. Hence, if we find a $satisfiable$ counter-example trace to the safety properties of all generators we return it as the desired FDIA, else we stop the verification with no possible FDIA vector (line \ref{algsyn:ifunsat}). Algo.~\ref{alg:synnval} returns the FDIA on finding a satisfiable solution to $\phi$ along with the accompanying LAA as a $d$-length attack vector $A_d$ that is verified to be successful and stealthy (line \ref{algsyn:ret}).

\section{Experimental Setup and Results}\label{secResult}
For our experimental evaluation, we consider a standard Simulink model of an IEEE-14 bus power system (Fig. \ref{fig:figtool}). 

For the implementation of the RL agent, we used the RL toolbox in Matlab. The training of the RL attacker agent is done using the simulated environment in Simulink. For training of the RL agent and the subsequent FDIA  synthesis using S-TaLiRo, we use a 16-core Intel Xeon CPU with 32 GB of RAM. To visualize the effects of the synthesized stealthy attack vectors in real-time, the IEEE-14 bus power system model is implemented and emulated along with the LAA and FDIA vectors on an Opal  RT HIL setup (OP4510) using Opal RT real-time  simulation engine. The repository for the tool framework can be found here\footnote{https://anonymous.4open.science/r/smart-grid-attack-synthesis-tool/}.
\begin{wrapfigure}[]{l}{0.558\columnwidth}
	\includegraphics[clip,width = 0.55\columnwidth]{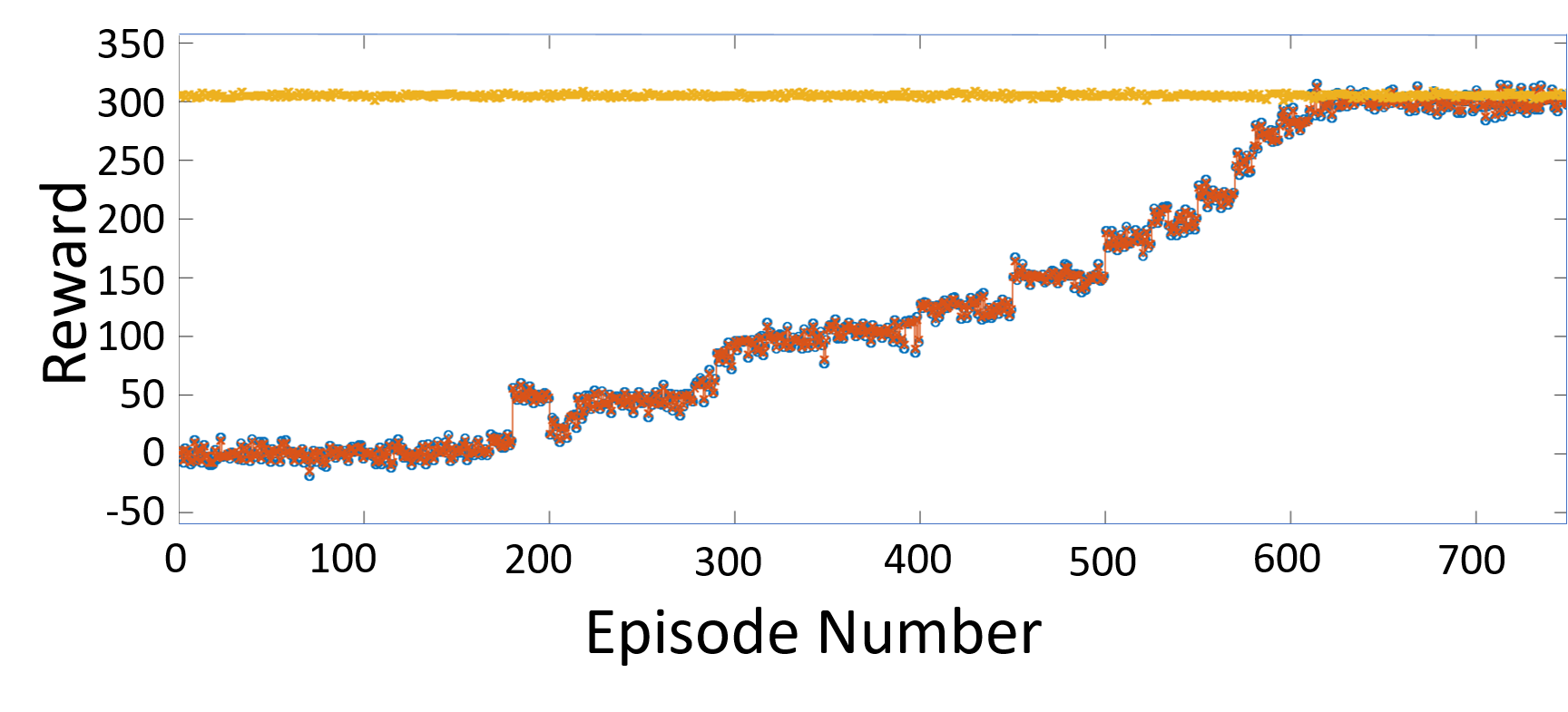}
	\centering
	\caption{RL agent training}
	\label{figRltrain}
 \end{wrapfigure}
As discussed in Sec.~\ref{secMeth} our framework takes the IEEE 14-bus power system Simulink model and trains the attacker RL agent to learn optimal LAA (Sec.~\ref{secAtkRl}). The 
attacker RL agent is designed and trained for 800 episodes, each spanning for 800 iterations or 8 seconds intervals. As we can see in Fig.~\ref{figRltrain}, the deployed DDPG agent is able to successfully maximize its reward during training and stabilizes at $300$ at the end of 800 episodes. The tool can generate multiple circuit breaker control signal sequences as instructed, that are able to launch LAA on the IEEE 14-bus power system model without raising any alarm. But as discussed earlier (Sec.~\ref{secAtkMod}) the instability induced by these LAA is not retained for long. In the next step, our tool chain takes these circuit breaker control signal sequences and models them in the power grid Simulink model to simulate the load alteration attack scenarios as explained in Algo.~\ref{alg:synnval}. Our  tool invokes S-TaLiRo with required inputs (as explained in Sec.~\ref{secFormalAtkSyn}). S-TaLiRo verifies and generates a vulnerable false data sequence for each LAA sequence, which can successfully push the system states  towards unsafe region within a short duration, without raising any alarm.
\begin{figure}[!htb]
\centering
\begin{subfigure}[b]{0.48\columnwidth}
\includegraphics[clip,width=1.05\textwidth]{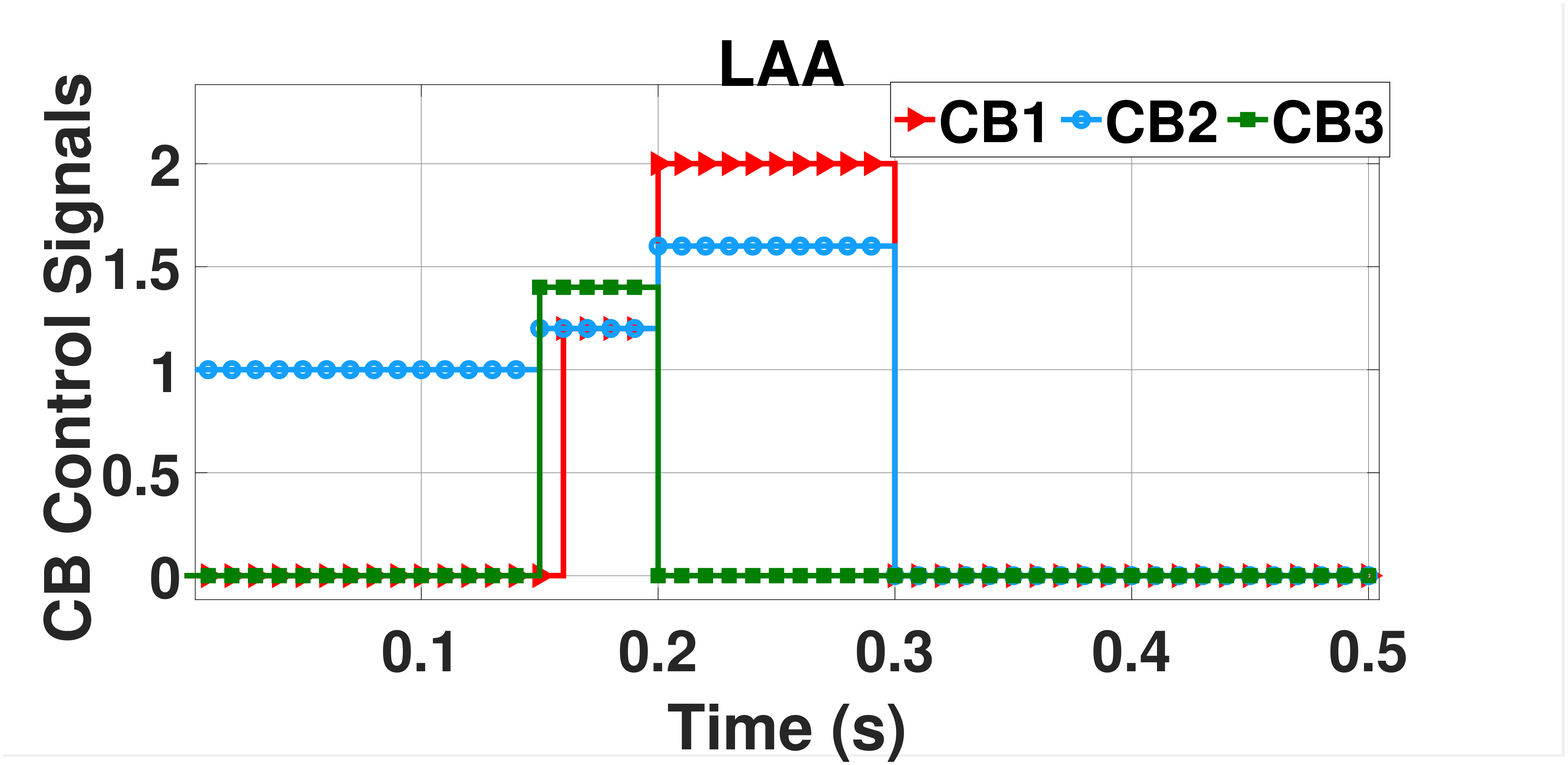}
  \subcaption{Circuit Breaker Signals}
  \label{fig:laa}
 \end{subfigure}
\begin{subfigure}[b]{0.49\columnwidth}
\includegraphics[clip,width=1\textwidth]{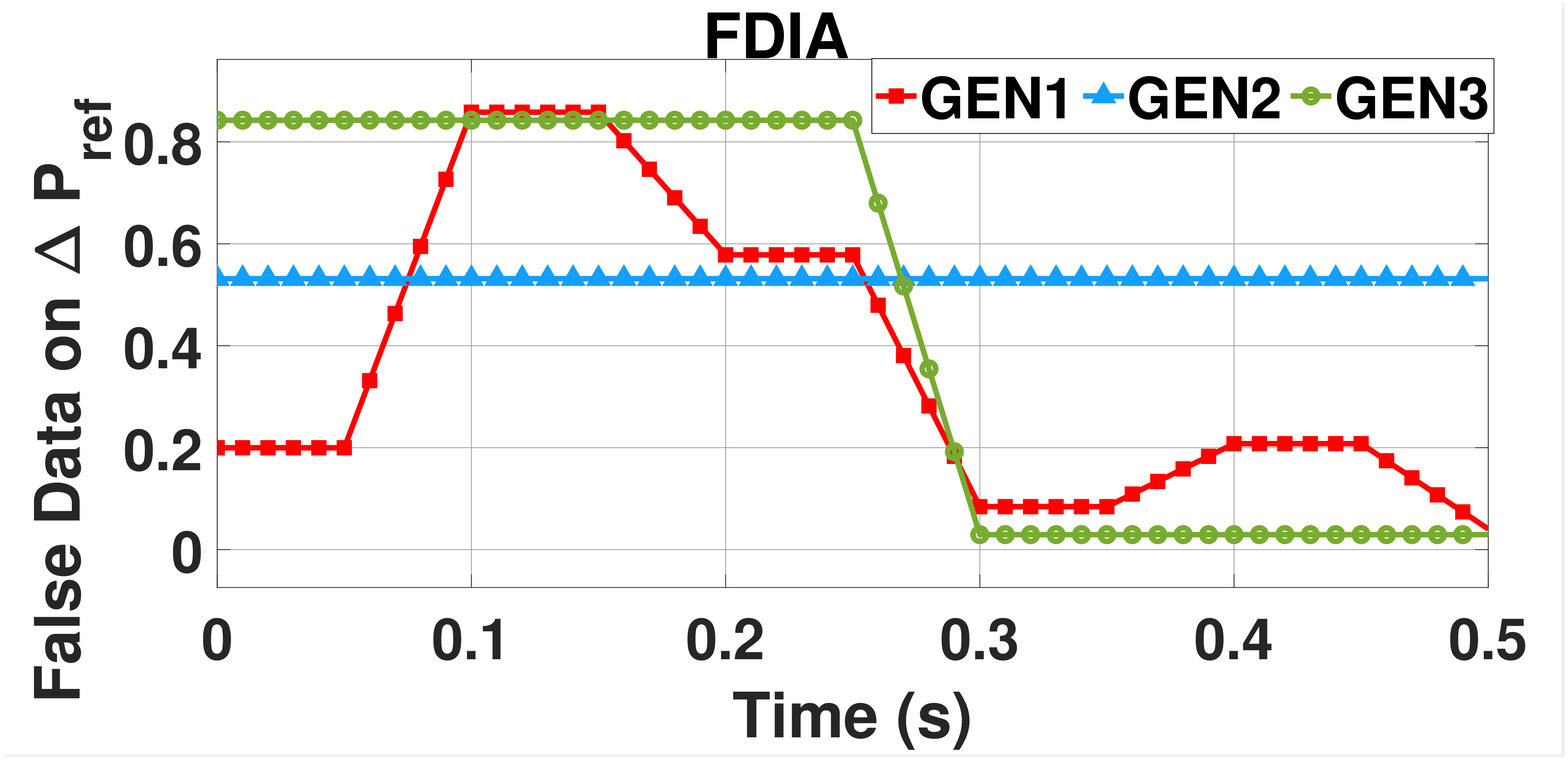}
  \subcaption{False Data Sequence on $\Delta P_{ref}$}
  \label{fig:fdia}
 \end{subfigure}%
\caption{LAA and FDIA Attack Vectors}
\label{fig:atksig}
\end{figure}

We demonstrate our results for one candidate output LAA+FDIA sequence as plotted in Fig.~\ref{fig:atksig}. 
The $red$ (with triangle markers), $blue$ (with circle markers) and $green$ (with square markers) plots in Fig.~\ref{fig:laa} denote the control signals for the circuit breakers $CB1$, $CB2$, and $CB3$ respectively  synthesized by our trained RL-based attacker.
The $red$ (with square markers), $blue$ (with triangle markers) and $green$ (with circle markers) plots in Fig.~\ref{fig:fdia} are false data sequences synthesized by S-TaLiRo as part of our tool-flow. They  are to be injected to the $\Delta P_{ref}$ outputs of the AGC loops in $GEN1$, $GEN2$, and $GEN3$ respectively.
We launch these synthesized attack vectors in the IEEE 14-bus model implemented in the OPAL-RT HIL and plot how it changes the usual behaviour of the power grid towards unsafe operations.
\begin{wrapfigure}{l}{0.5\columnwidth}
\centering
\includegraphics[clip,width=0.53\columnwidth]{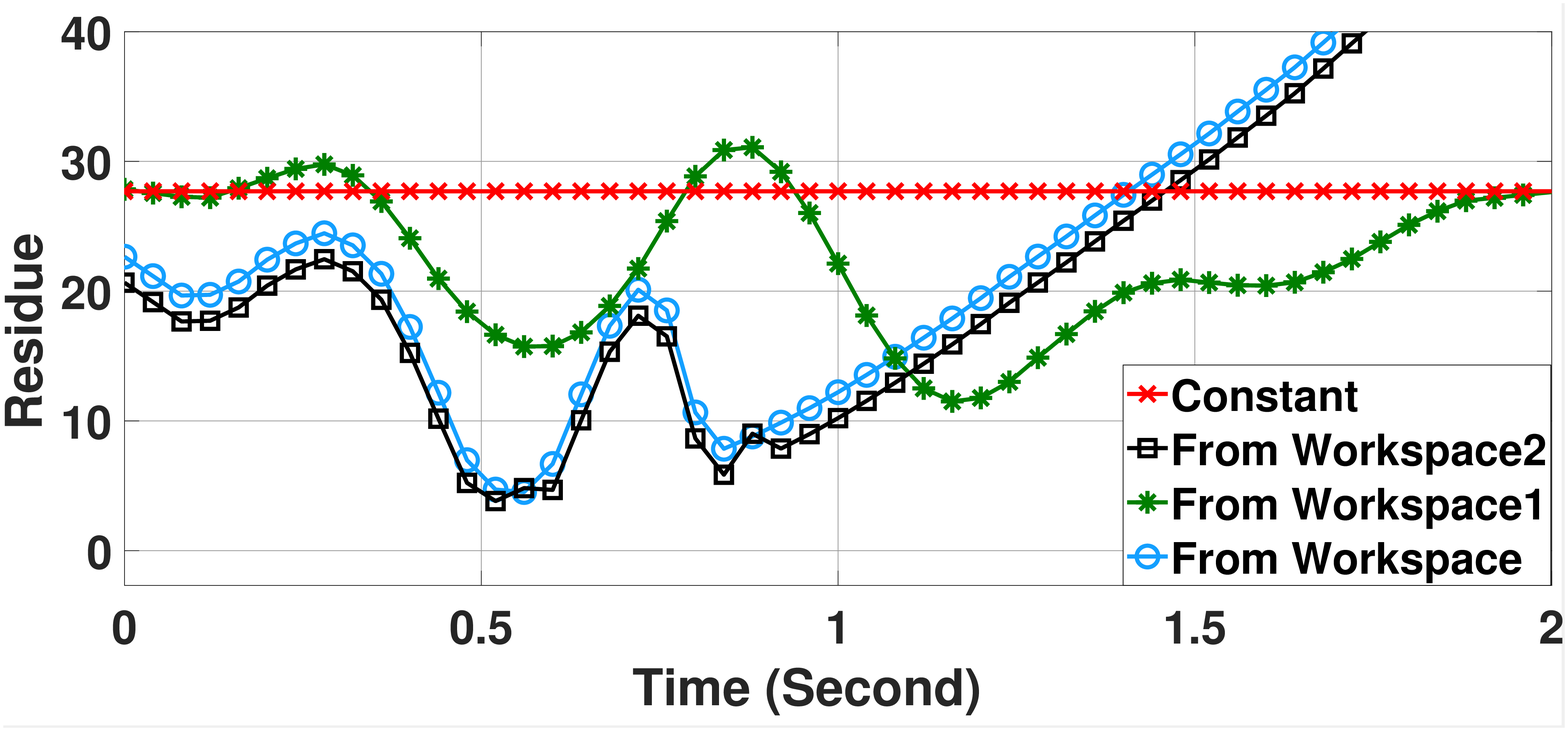}
\caption{Residue under attack and no-attack in HIL}
\label{fig:figcompres}
\end{wrapfigure}
In Fig.~\ref{fig:figcompres}, the $red$ plot with diamond markers denotes the nominal residue of the AGC loop in $GEN1$ under no attack situation and the $blue$ plot with square markers denotes the residue of the same under the synthesized LAA and FDIA. We consider the nominal residue during the normal operation of the grid as the threshold ($Th$). It can be observed that the residue under attack 
goes beyond the threshold ($Th$). Therefore the proposed attack strategy gets detected by the anomaly detection unit within $0.2$ seconds (i.e. $20$ sampling iterations for the AGC loop). Now let us observe how the synthesized LAA+FDIA affects the power grid model within 0.2 seconds period, i.e. before it gets detected.


\begin{figure*}[!t]
    \begin{subfigure}{0.65\columnwidth}
        \includegraphics[width=\linewidth]{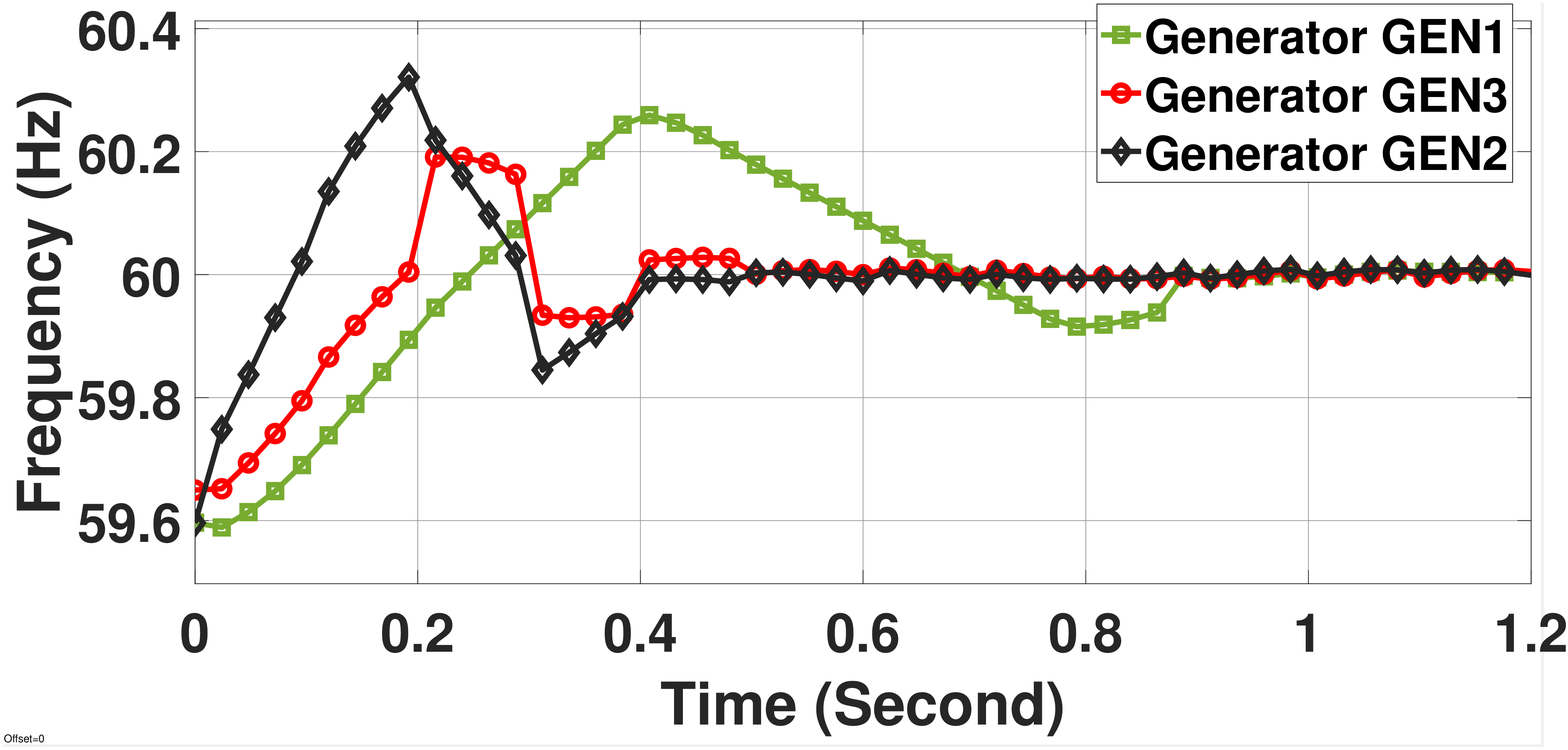}
  \caption{Generator frequency in normal condition}
  \label{fig:freqNoAttack}
    \end{subfigure}
    \hfill
    \begin{subfigure}{0.65\columnwidth}
        \includegraphics[width=1\textwidth]{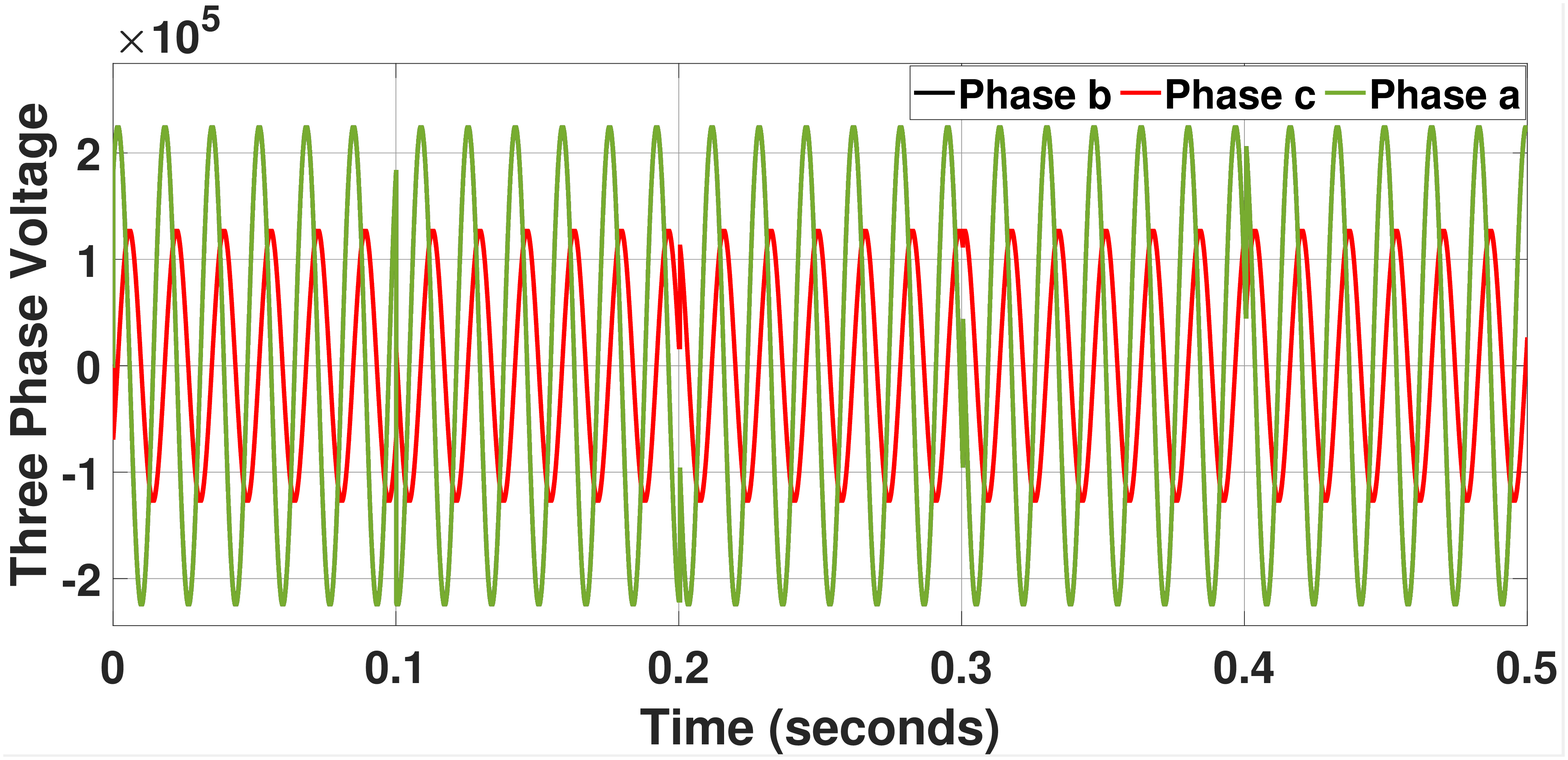}
  \caption{Transmission line voltage in normal condition}
  \label{fig:volNoAttack}
    \end{subfigure}
    \hfill
    \begin{subfigure}{0.65\columnwidth}
        \includegraphics[width=1\textwidth]{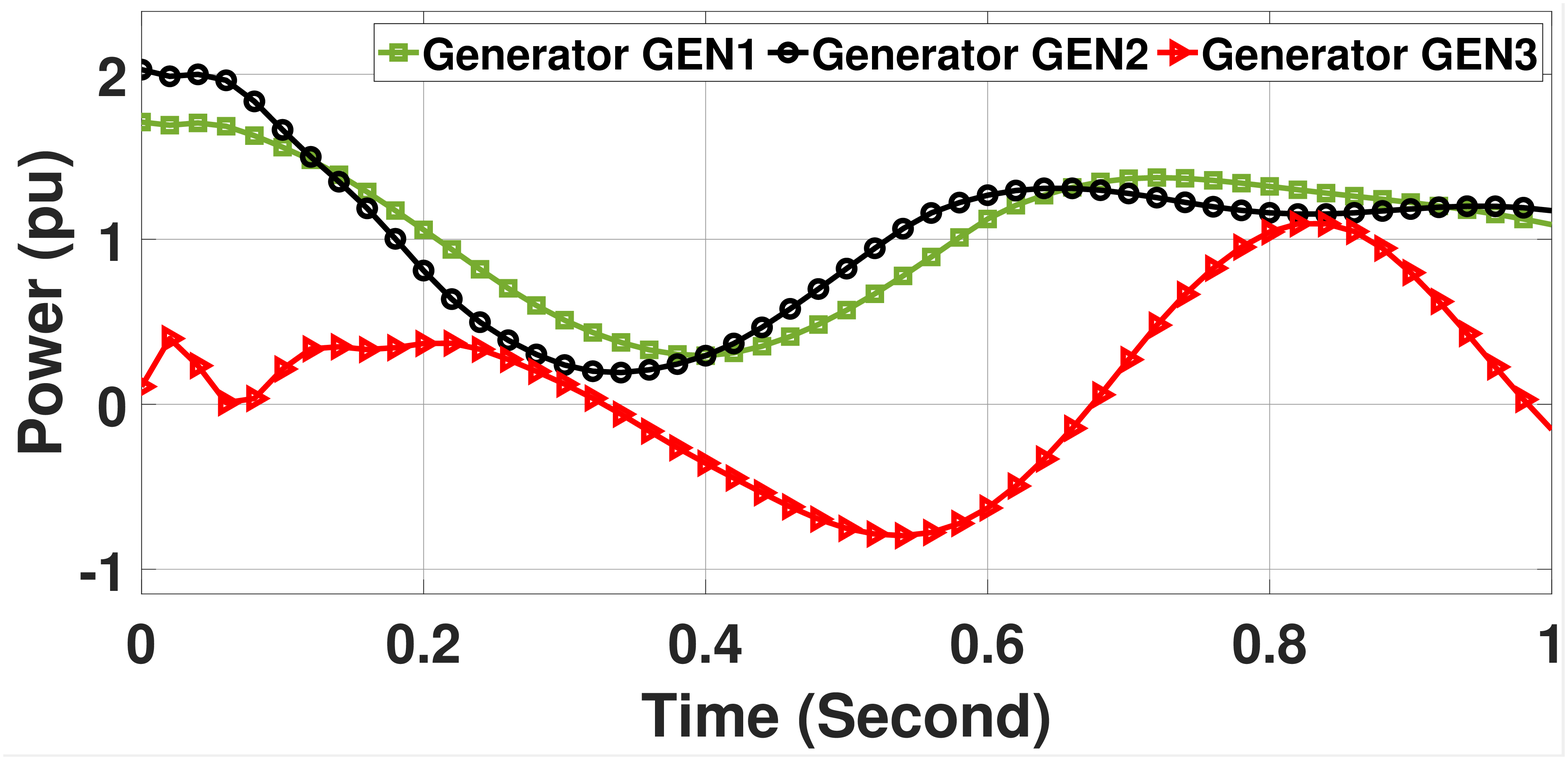}
	\centering
	\caption{Generator output electrical power in normal condition}
  \label{fig:powerNoAttack}
    \end{subfigure}
    \begin{subfigure}{0.65\columnwidth}
        \includegraphics[width=\linewidth]{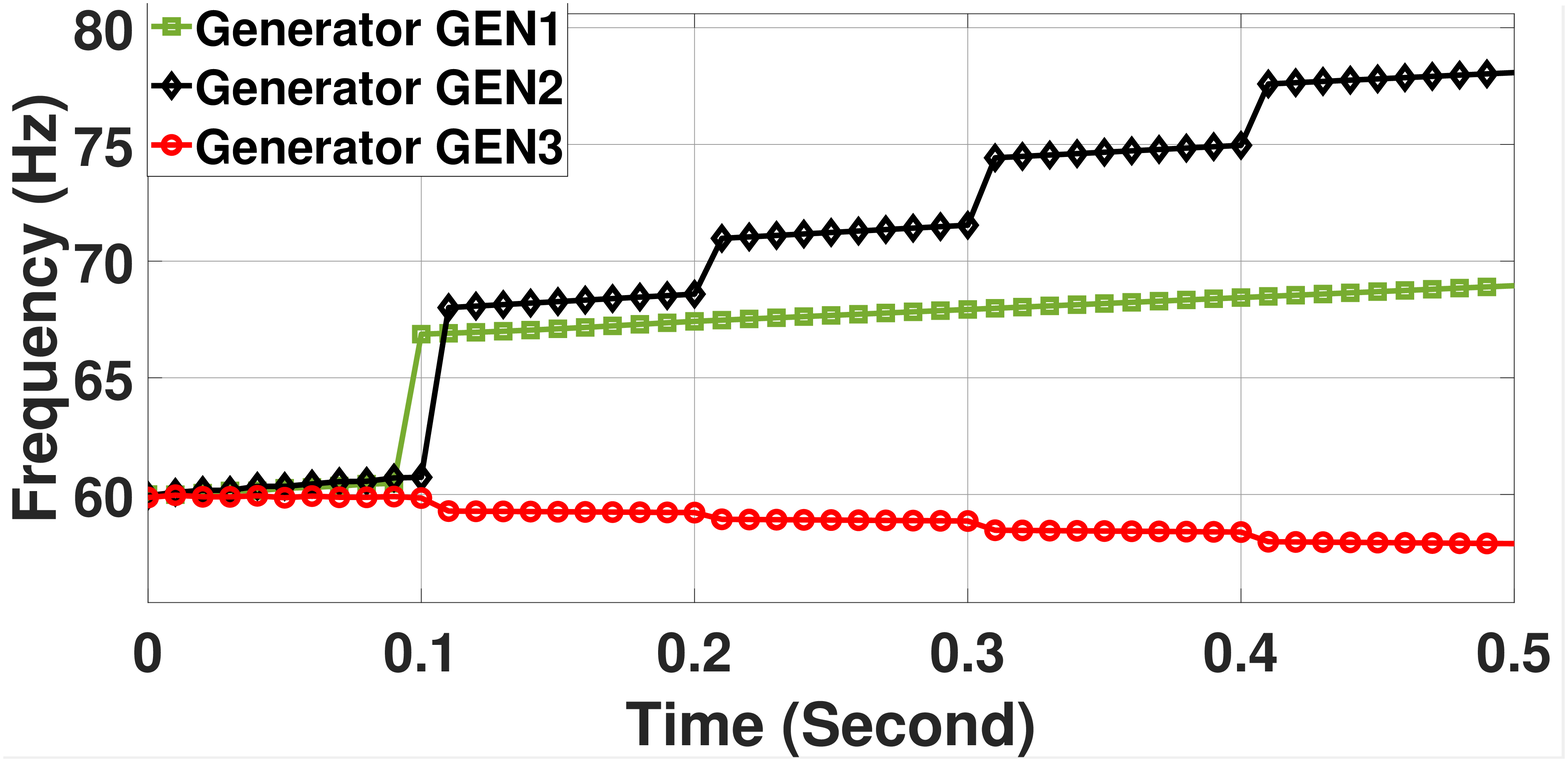}
  \caption{Generator frequency under attack}
  \label{fig:freqAttack}
    \end{subfigure}
    \hfill
    \begin{subfigure}{0.65\columnwidth}
        \includegraphics[width=1\textwidth]{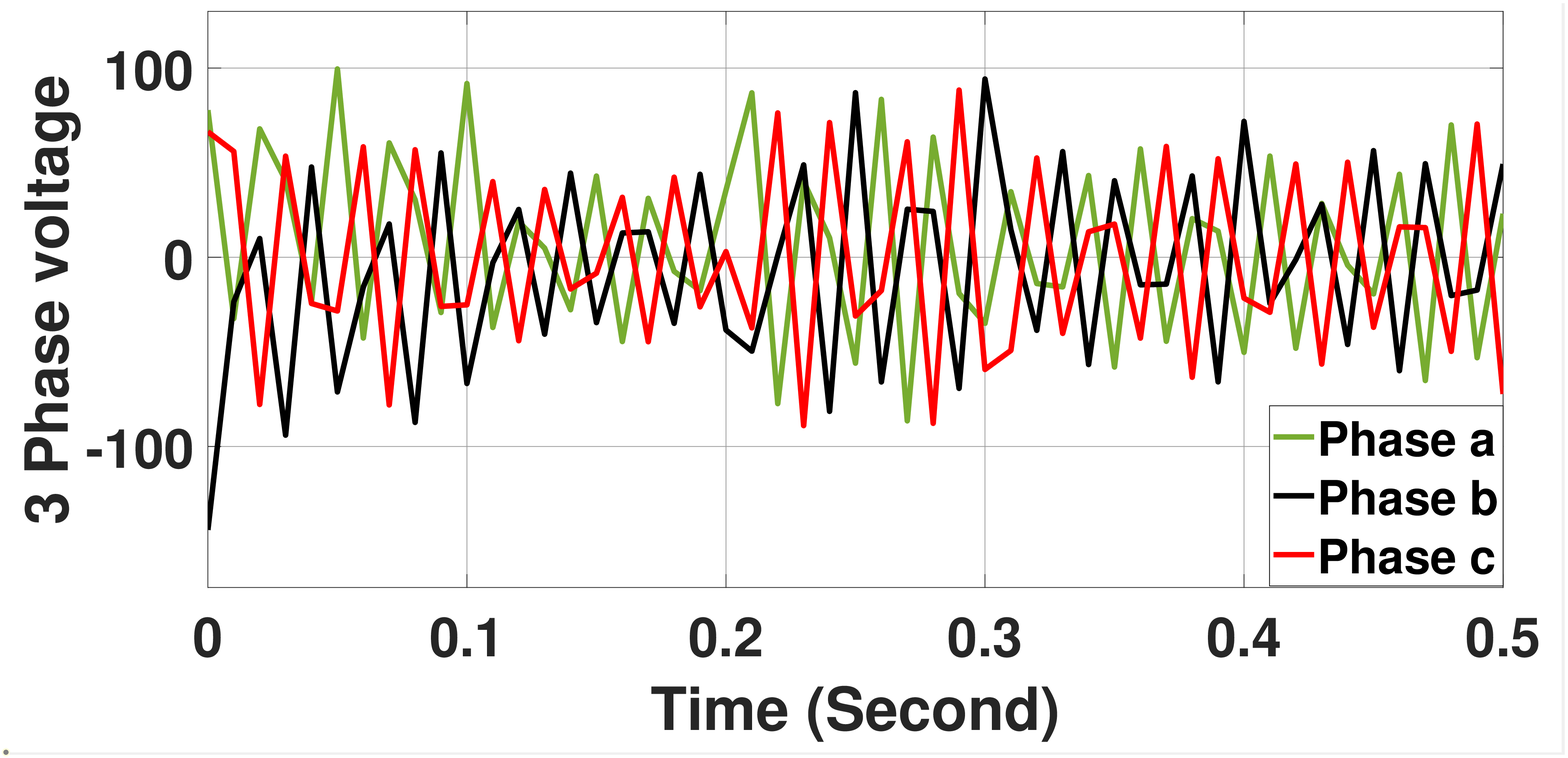}
  \caption{Transmission line voltage under attack}
  \label{fig:volAttack}
    \end{subfigure}
    \hfill
    \begin{subfigure}{0.65\columnwidth}
        \includegraphics[width=1\textwidth]{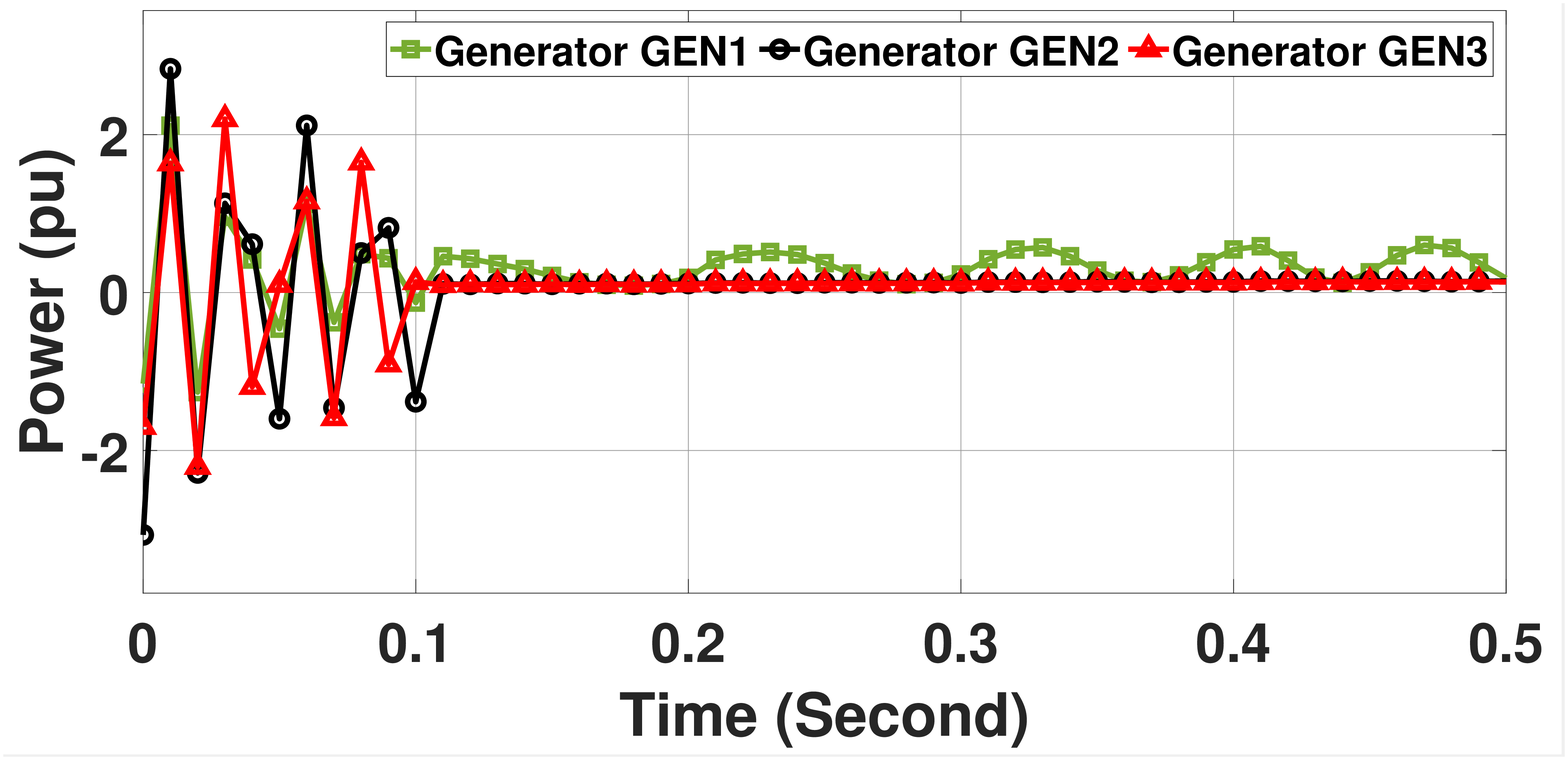}
	\centering
	\caption{Generator output electrical power under attack}
  \label{fig:powerAttack}
    \end{subfigure}
 \caption{Simulation result in HIL setup}
 \vspace{-4 mm}
 \label{figAttackSimulationHIL}
\end{figure*}
We visualize the effect of the proposed attack on the grid by plotting the following grid parameters in Fig.~\ref{figAttackSimulationHIL}. 
\textit{(i)} \textbf{Operating frequency of the generation units: }  
In Fig.~\ref{fig:freqNoAttack} we plot the operating frequencies of generator units during normal operation, and in Fig.~\ref{fig:freqAttack} we plot the same under synthesized attacks. The $green$ (with square markers), $black$ (with diamond markers), and $red$ (with circle markers)  denote the operating frequencies of $GEN1$ ($f^1$), $GEN2$ ($f^2$) and $GEN3$ ($f^3$) respectively. As can be seen in Fig.~\ref{fig:freqNoAttack}, during normal operation the frequencies stay within their desired safe region of operation i.e. $[59.5,60.5]$, and stabilize eventually. When the synthesized attack vectors are in play, the frequencies under attack rapidly increase and go beyond the safety boundary within 0.2 seconds, i.e. before the attack gets detected. Moreover, the generation units lose synchronism thereafter as the frequencies never return back to the steady state. \textit{(ii)} \textbf{Voltage of  the transmission line} connecting bus 2 and 3 (refer to Fig~\ref{fig:figtool}): Fig.~\ref{fig:volNoAttack} and Fig.~\ref{fig:volAttack} plots the voltage of the transmission line without attack and under attack respectively. The green, black and red plots denote 3 phase voltages (V). As can be seen in Fig.~\ref{fig:volNoAttack} under normal operation 3 phase voltage of the transmission line follows a uniform  sinusoidal waveform with magnitude in the order of $10^5$. Whereas under attack the 3 phase voltage profile shows a distorted waveform with magnitude in the order of $10$ which causes brownouts and eventually leads to a permanent blackout.
\textit{(iii)} \textbf{Output electrical power of synchronous generator units:} In Fig.\ref{fig:powerNoAttack} and Fig.\ref{fig:powerAttack} we plot the electrical power outputs of the generator units without attack and under attack situations respectively. The green (with square marker), black (with circle marker), and red (with triangle marker) plots are the power outputs of $GEN1$($P_e^1$), $GEN2$($P_e^2$) and $GEN3$($P_e^3$) respectively. As can be seen, under a no-attack situation the output power waves stabilize to a fixed value (for $GEN1$,$GEN2$) or change as per the demand ($GEN3$). But under attack, the power outputs become zero within 0.2 seconds or oscillate uncontrollably due to its raised frequency ($GEN1$).

\par Therefore, the attack vectors synthesized using our tool are able to sense the difference in desired power reference $\Delta P_{ref}$ from the change in grid frequency and accordingly update the $P_{ref}$ to severely damage the normal operation of a power grid while being stealthy. By the time the attack gets detected by the residue-based detector, the system faces a blackout. 
Moreover, our detection unit is responsive to the transient characteristics of the AGC as well. The designed framework is able to scalably explore the multi-dimensional attack surface and output learning-based optimal LAA, combined with FDIA, synthesized using a stochastic falsification engine. Visualizing the effects of the synthesized attack vectors in presence of such a detection unit in real-time implementation of an IEEE 14-bus power system validates the efficacy of our framework.
\section{Conclusion}
This paper presents a novel RL-based framework for the synthesis of \emph{input load alteration} and \emph{output data falsification} attack vectors that are verified to make an input power grid model unsafe without being detected. The framework scalably explores the multi-dimensional attack surface of any complex power grid model by utilizing both learning-based and simulation-based falsification engines. As a future extension, we plan to incorporate the probabilistic analysis of the learned LAA sequences to quantify their success rate when combined with the synthesized FDIA. We also plan to design learning-enabled verifiable detection and mitigation strategies to counter such attacks.

\bibliographystyle{ieeetr}
\bibliography{Ref}
\end{document}